\documentclass[a4paper, 5p, sort&compress]{elsarticle}

\usepackage{graphicx}
\usepackage{amssymb}
\usepackage{amsmath}

\usepackage{hyperref}

\usepackage{url}

\journal{Elsevier}

\begin{document}

\begin{frontmatter}


\title{Counter-intuitive behaviour of energy system models under CO$_2$ caps and prices}



\author[fzj,col]{Juliane Weber}
\author[fzj]{Heidi Ursula Heinrichs}
\author[fzj]{Bastian Gillessen}
\author[fzj]{Diana Schumann}
\author[fias]{Jonas H\"orsch}
\author[fias,kit]{Tom Brown}
\author[fzj,col]{Dirk Witthaut}
\ead{d.witthaut@fz-juelich.de}

\address[fzj]{Forschungszentrum J\"ulich, Institute for Energy and Climate Research -- 
              Systems Analysis and Technology Evaluation, 52428 J\"ulich, Germany}
\address[col]{University of Cologne, Institute for Theoretical Physics, Z\"ulpicher Str. 77,
               50937 Cologne, Germany}
\address[fias]{Frankfurt Institute for Advanced Studies, 60438 Frankfurt am Main, Germany}
\address[kit]{Karlsruhe Institute of Technology, Institute for Automation and Applied Informatics, 76344 Eggenstein, Germany}

\begin{abstract}
The mitigation of climate change requires a fundamental transition of the energy system. Affordability, reliability and the reduction of greenhouse gas emissions constitute central but often conflicting targets for this energy transition. Against this context, we reveal limitations and counter-intuitive results in the model-based optimization of energy systems, which are often applied for policy advice. When system costs are minimized in the presence of a CO$_2$ cap, efficiency gains free a part of the CO$_2$ cap, allowing cheap technologies to replace expensive low-emission technologies. Even more striking results are observed in a setup where emissions are minimized in the presence of a budget constraint. Increasing CO$_2$ prices can oust clean, but expensive technologies out of the system, and eventually lead to higher emissions. These effects robustly occur in models of different scope and complexity. Hence, extreme care is necessary in the application of energy system optimization models to avoid misleading policy advice.
\end{abstract}

\begin{keyword}
energy system model \sep optimization \sep CO$_2$ cap \sep CO$_2$ tax
\end{keyword}

\end{frontmatter}


\section{Introduction}

The mitigation of climate change requires a fundamental transition of the energy system. Currently, 65\% of all greenhouse gas emissions are caused by the carbon dioxide (CO$_3$) emissions from fossil fuel combustion and industrial processes \cite{ipcc2014}, such that a rapid decarbonisation of the energy sector is inevitable to meet the 2$^\circ$C goal of the Paris agreement \cite{Paris15,Roge15,Rogelj16,Rockstrom2017,Rogelj2017}. Fossil fuelled power plants must be replaced by renewable sources such as wind turbines and solar photovoltaics, whose costs are becoming more and more competitive \cite{Jacobson2011,Delucchi2011,chu12,Creutzig2017}. One of the largest challenges of this transition concerns the security and reliability of the energy supply, which is crucial for industry, economy and infrastructure operation \cite{Sims11,Brummitt2013,15focus,18cascade} as well as the public acceptance of the transition \cite{Sovacool2016}. Wind and solar power generation are inherently fluctuating \cite{Milan2013,Olauson2016,17fluctuations}, and suitable locations are often far away from the centers of the load \cite{georgilakis2008,rodriguez2014,18redispatch}. The design of a future energy system must respect these constraints to guarantee a sustainable and reliable supply at affordable costs \cite{Roge15,chu12,IEA17b,Walsh2017}.

\begin{figure}[tb]
\centering
\includegraphics[clip=true, trim= 4.2cm 8cm 6.6cm 1cm, width=\columnwidth, angle=0]{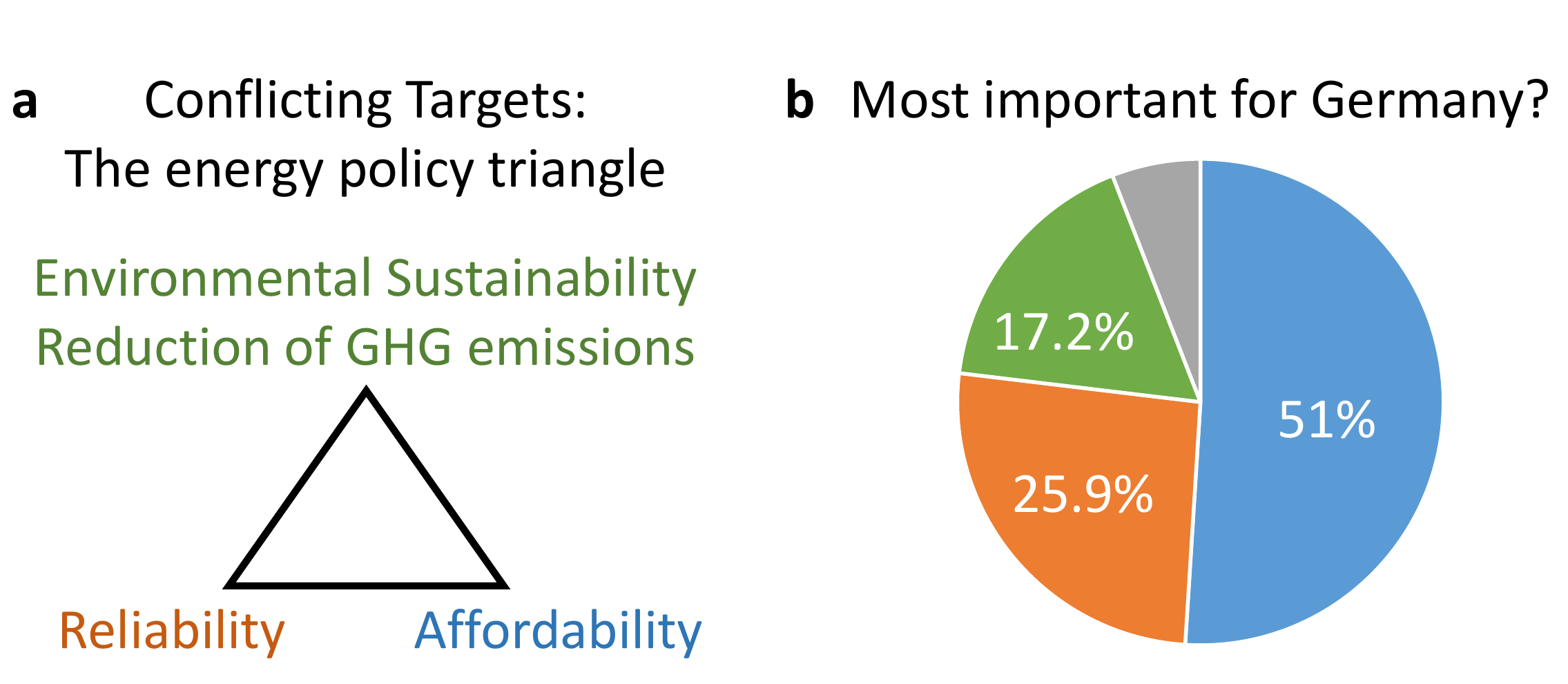}
\caption{\label{fig:triangle}
{Conflicting targets in energy system optimization and planning.}
(a) Energy policy triangle consisting of the targets affordability, reliability and the reduction of greenhouse gas (GHG) emissions.
(b)
In a representative survey, 1006 people in Germany were asked to rank five different aspects of energy security by their national importance. The aspect ``Affordability of Electricity and Heat" was ranked highest by 51\% of the participants (blue), but the aspects ``Reliable Supply with Oil, Gas and Other Energy Carriers" (red, 25.9\%) and ``Reduction of GHG emissions" (green, 17.2\%) were also named as first priority frequently. Results from an own panel survey with 1006 respondents, carried out in 2014 \cite{Schumann2017}.
}
\end{figure}

Affordability, reliability and environmental sustainability constitute central targets for energy policy, with the reduction of greenhouse gas (GHG) emissions being the most urgent environmental target (Fig.~\ref{fig:triangle}a). This set of targets is commonly referred to as the energy policy triangle. It forms the basis for the energy strategy of the European Union \cite{EU12,EU2015} and is widely supported by the public. A representative survey in Germany shows that half of the population ranks affordability as the most important goal, but reliability and reduction of GHG emissions are also named as first priority frequently (Fig.~\ref{fig:triangle}b). However, the three targets are often conflicting, so that the triangle becomes a trilemma \cite{WEC2016}. None of these targets can be abandoned or singled out to the exclusion of the others. As a result, balancing the targets and resolving conflicts between them is at the heart of energy system analysis and energy policy.

A variety of approaches has been put forward to assess and optimize energy systems based on these targets. Modelling approaches range from purely technical through techno-economic to predominantly economic models and most recently socio-technical models \cite{PFENNINGER201474,JEBARAJ2006281, 17intESM}. They differ in scales, system boundaries and level of detail. Among these modelling approaches,  the class of techno-economic optimization under constraints is particularly wide-spread \cite{Loulou2008,Howells2011,HEINRICHS2017234,JRC-TIMES, PFENNINGER201474,JEBARAJ2006281}. Conflicting targets can be integrated using a specific weighting scheme or via constrained optimization. The results of such models are often fed directly into the political decision making process.

Against this context, we reveal limitations and counter-intuitive results in the techno-economic optimization of energy systems. We show that in a common emission-constrained cost optimizing model, the improvement of a technology can impede its utilization -- an effect that may discourage innovations and investments. Even more striking, emission minimization in the presence of a budget limit can lead to  effects reminiscent of Giffen's paradox in microeconomics \cite{Stigler1947,Heijman2002}. In such a context, the increase of effective CO$_2$ costs can lead to higher CO$_2$ emissions. 

We illustrate these findings for three energy system optimization models of different scope and complexity. We first consider an elementary model to reveal fundamental interactions of different constraints and objectives, then we show that effects manifest both in a short-term electricity sector model and a long-term integrated energy system model and draw some key conclusions.

\section{Methods}

In this article, we analyze the effects caused by target conflicts for three different energy system optimization models. We first consider a very stylized model to introduce the basic phenomena and then consider a detailed electricity sector model and an integrated energy system model. All types of models operationalize the three central targets of energy policy as follows:
\begin{itemize}
    \item Affordability: Reduce total system costs $C$,
    \item Sustainability: Reduce total GHG emissions $E$,
    \item Reliability: Satisfaction of demand (all models) plus model specific constraints such as power grid stability.
\end{itemize}
In an energy system optimization model one of the first two targets is promoted to the objective function which is minimized, while the other targets are included via constraints. 

We consider different combinations of objectives and constraints to explore fundamental problems arising from conflicting targets. In addition to the generic cost minimization, we also explore a hypothetical setting where emissions are minimized. Hence, we obtain the two cases:
\begin{itemize}
    \item Case A: Minimization of total system costs $C$ with a hard emission cap and reliability constraints
    \item Case B: Minimization of total emissions $E$ with a budget cap and reliability constraints
\end{itemize}

\subsection{Elementary Model}
\label{sec:el-mod}

We first consider an elementary decision problem, including only two fossil fuels used for electricity generation: one cheap type with high specific CO$_2$ emissions (e.g. lignite) and one expensive type with low specific CO$_2$ emissions (e.g. natural gas). The model then optimizes the energy mix, i.e. the total electricity generation $G$ from the two fuel types per period.

\begingroup
\begin{center}
\setlength{\tabcolsep}{4pt} 
\renewcommand{\arraystretch}{1.5}
\setlength{\doublerulesep}{1.0pt}
\begin{table}[b]
\caption{Input parameters to derive the variable costs. Shown are the fuel costs using the lower heating value ($vc_{\rm fc,lhw}$), the net efficiency ($\eta$), the CO$_2$ certificate prices ($c_{\rm CO_2}$), the specific emissions per fuel ($se_f$) and the variable costs for maintenance ($vc_{\rm maintenance}$) and for operating materials ($vc_{\rm materials}$). Values taken from \cite{Kons13}.}
\label{tab:vc}
\begin{tabular}{lrrr}
& Unit & Lignite & Gas \\
\hline
$vc_{\rm fc,lhw}$ & Euro/MWh & 5.40 & 36.30 \\
$\eta$ & \% & 45 & 56 (case B) \\
$c_{\rm CO_2}$ & Euro/t & 15 (case A) & 15 (case A) \\
$se_f$ & t/MWh$_{\rm th}$ & 0.41 & 0.202 \\
$vc_{\rm maintenance}$ & Euro/MWh & - & 3 \\
$vc_{\rm materials}$ & Euro/MWh & 1.65 & 0.5 \\
\hline
\end{tabular}
\end{table}
\end{center}
\endgroup

The model implements the reliability target via a hard constraint: The electricity demand $D$ per period must always be satisfied, such that we have the inequality \begin{equation}
    G_{\rm gas} + G_{\rm lignite} \ge D.
    \label{eq:demand-constraint}
\end{equation}
Furthermore, the model seeks to optimize the total system costs and total CO$_2$ emissions:
\begin{align}
      C &= vc_{\rm lignite} \times G_{\rm lignite} + vc_{\rm gas} \times G_{\rm gas}, \label{eq:cost-constraint} \\
      E &= se_{\rm lignite} \times G_{\rm lignite} + se_{\rm gas} \times G_{\rm gas}, \label{eq:emission-constraint}
\end{align}
where $vc_{\rm fuel}$ denotes the variable costs and $se_{\rm fuel}$ the specific CO$_2$ emissions for the two fuel types. Throughout this paper, we focus on CO$_2$ emissions, as these are the main contributor to climate change. Other greenhouse gases can readily be included in terms of CO$_2$ equivalents. We either optimize the costs $C$ in the presence of an emission cap (case A) or the emissions $E$ in term of a budget cap (case B). 

The parameters used in our study are calculated as follows. The variable costs consist of costs for fuel (fc), CO$_2$ emissions, maintenance and operating materials:
\begin{equation}
  vc = vc_{\rm fc} + vc_{\rm CO_2} + vc_{\rm maintenance} + vc_{\rm materials} .
\end{equation}
Fuel costs are given using the lower heating value (lhw). They thus depend on the net efficiency $\eta$, which is varied for gas in case A:
\begin{equation}
  vc_{\rm fc} = \frac{vc_{\rm fc,lhw}}{\eta}.
\end{equation}
The costs for CO$_2$ emissions depend on the CO$_2$ certificate price $c_{\rm CO_2}$, which we vary in case B, and on the emissions per generated MWh of electricity:
\begin{equation}
   vc_{\rm CO_2} =  \frac{se_f}{\eta} \cdot c_{\rm CO_2}.
\end{equation}
with $se = se_f/\eta$ being the electricity specific emissions. The single parts of the variable costs are derived from the values listed in Table~7.12 and Example~7.4 in \cite{Kons13} and are summarized in Table \ref{tab:vc}.

We assume a fixed installed capacity for each power plant type and choose it such that one type can meet the demand without the other. Hence, if $G_{\rm gas}$ = 0, then $G_{\rm lignite} = D$ and vice versa.

The constraint is chosen such that the optimization problem is always solvable and non-trivial. Thus, in case A we choose $E_{\rm cap}$ equal to the maximum specific emissions $se$ of gas:
\begin{equation}
  E_{\rm cap} = \max[se_{\rm gas}] \times D = \frac{se_{f,\rm gas}}{\min[\eta_{\rm gas}]} \times D.
\end{equation}
A higher value would allow for more lignite in the system and $E_{\rm cap} \geq se_{\rm lignite} \times D$ would lead to the trivial solution with only lignite being used for all $\eta_{\rm gas}$. Similarly, the CO$_2$ price is chosen to be low enough to avoid gas being substituted for coal on a cost basis, which would lead to the trivial solution with only gas being used in all cases.

In case B, we set $C_{\rm budget}$ equal to the minimum variable costs of the highest CO$_2$ certificate price, i.e.
\begin{equation}
\begin{aligned}
C_{\rm budget} = \min[vc_{\rm lignite}(\max[C_{\rm CO_2}]) & \times D,\\
vc_{\rm gas}(\max[C_{\rm CO_2}])& \times D].
\end{aligned}
\end{equation}
A higher budget constraint would allow for more gas in the system and choosing $C_{\rm budget} \geq vc_{\rm gas}(\max[C_{\rm CO_2}]) \times D$ would lead to the trivial solution that only gas is used for all CO$_2$ certificate prices. Similarly, a sufficiently high CO$_2$ price would also cause coal to be replaced by gas, provided that the budget would be high enough for this solution to be feasible.

\subsection{Electricity System Model PyPSA}
\label{sec:pypsa}

\begin{figure*}[tb]
\centering
\includegraphics[width=15cm, angle=0]{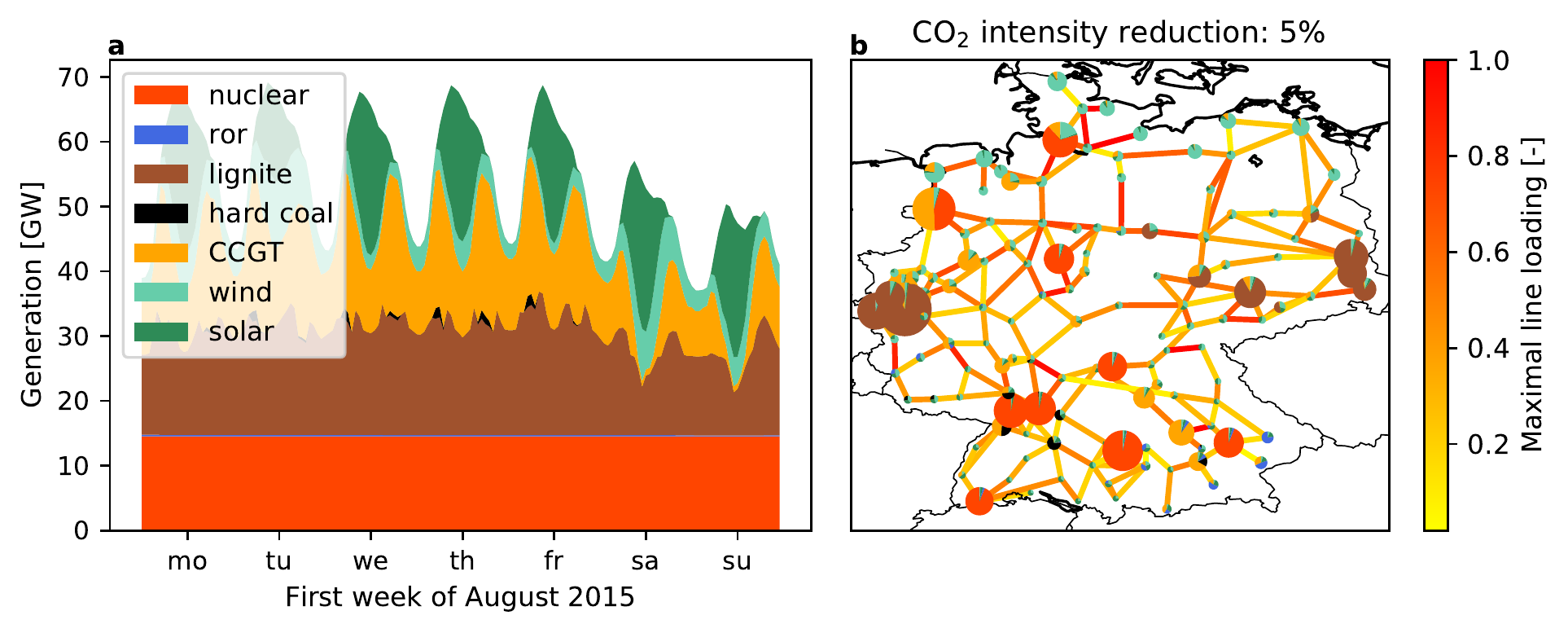}
\caption{\label{fig:pypsa}
{Optimization in a detailed electricity sector model.}
We analyse a model of the German electricity sector with high spatial
and temporal resolution.
(a) The model optimizes the dispatch of generators and storage facilities as well as the curtailment with hourly resolution over a full year. The figure shows the resulting operation for one exemplary weak in August.
(b) The model takes into account a variety of reliability constraints. Most importantly, the demand must be satisfied for all nodes and no transmission line may be overloaded. 
The color code shows the maximum relative line loading during the year, which may not exceed one. The pie charts show the annually aggregated generation for every node and primary energy carrier.
CCGT: combined cycle gas turbines, ror: run-of-river.
}
\end{figure*}

The electricity sector model PyPSA optimizes the operation of a representation of the German power system for the year 2015 with a high spatial and temporal resolution. The model includes conventional and renewable power generators, pumped hydro storage units, transmission lines and the electrical demand. The dispatch of power plants and storage as well as a potential curtailment of renewable sources is optimised hourly for the full year using nodal pricing, guaranteeing that no transmission lines are overloaded and thus approximating the current system after market clearing and redispatch (but excluding energy trading with neighbouring countries). An exemplary optimization result for one weak is shown in \ref{fig:pypsa}a.

\begingroup
\setlength{\tabcolsep}{1.5pt} 
\renewcommand{\arraystretch}{1.7}
\setlength{\doublerulesep}{1.2pt}
\begin{table}[tb]
\caption{Input parameters to derive the variable costs in the electricity system model PyPSA. Shown are the fuel costs using the lower heating value ($vc_{\rm fc,lhw}$), the net efficiency ($\eta$), the CO$_2$ certificate prices ($c_{\rm CO_2}$), the specific emissions per fuel ($se_f$) and the variable costs for maintenance ($vc_{\rm maintenance}$) and for operating materials ($vc_{\rm materials}$). Values taken from \cite{schroeder2013}.
}
\label{tab:pypsa_costs}
\begin{tabular}{lrrrr r r}
& Unit & Nuclear & CCGT & OCGT & Hard Coal & Lignite\\
\hline
$vc_{\rm fc,lhw}$ & $\frac{\rm Euro}{{\rm MWh}_{\rm th}}$ & 3.0   & 21.6  & 21.6  & 8.4   & 2.9\\
$vc_{\rm om}$     & $\frac{\rm Euro}{{\rm MWh}}$          & 10.0  & 4.0   & 3.0   & 6.0   & 7.0\\
$\eta$            & \%                         & 33.7  & 61.0  & 39.0  & 46.4  & 44.7\\
$se_f$            & $\frac{\rm t}{{\rm MWh}_{\rm th}}$    & 0.000 & 0.181 & 0.181 & 0.336 & 0.333\\
\hline
\end{tabular}
\end{table}
\endgroup

The power system data corresponds to the German part of the European model PyPSA-Eur \cite{pypsa-eur}, implemented in the PyPSA modelling framework \cite{pypsa}. The software and all data are freely available online \cite{pypsa-website}. The hourly demand profiles are taken from the European Network of Transmission System Operators for Electricity (ENTSO-E) website \cite{entsoe_load}; the power plant database comes from the Open Power System Data (OPSD) project \cite{opsd-conventionalpowerplants}; the transmission grid data are based on the ENTSO-E interactive map \cite{interactive} extracted by the GridKit toolkit~\cite{wiegmans_2016_55853} and then clustered down to 128 major substations following the methodology in \cite{hoersch2017-eem}; the generation of time series for wind and solar power uses the methodology from \cite{REatlas}. The fuel costs, efficiencies and variable operation and maintenance (VOM) costs for conventional power plants are taken from \cite{schroeder2013} and listed in Table \ref{tab:pypsa_costs}. VOM costs for CCGT and OCGT refer to new installations, which are significantly lower than for older plants in the existing generation fleet. Solar, onshore and offshore wind and run-of-river are assumed to have zero variable costs. 

Several reliability constraints are implemented: in each time step the demand at each substation must be satisfied and transmission lines may not be overloaded (cf.~figure \ref{fig:pypsa}b). To approximate the $n-1$ network security constraint, it was enforced that no transmission line was ever loaded above 70\% of its thermal rating.

For case A, the CO$_2$ emissions cap was set by taking a 40\% reduction in emissions compared to the unconstrained cost minimum. This yields a cap of 112~Mt for the year. We study a technological development, that results in a reduction in CO$_2$ intensity for each technology, where a reduction of 0.1 results in the CO$_2$ emissions being reduced by 10\%. It is assumed that the efficiency is unchanged in order to isolate the effects of the reduction in carbon intensity on the model. Such a reduction of emissions can be realized with Carbon Capture and Storage (CCS). A reduction by 10\% would then correspond to the average over the entire generation fleet of one type, where reductions for different plants may well vary. CCS typically leads to a reduction of efficiency, which is not considered as described above.

The fixed budget for case B was derived by minimizing the system costs within the model assuming a CO$_2$ price of 50~Euro/t. Hence, no feasible solution can be found above 50~Euro/t.

\subsection{Energy System Model IKARUS}
\label{sec:ikarus}

\begin{figure}[bt]
\centering
\includegraphics[width=\columnwidth, angle=0]{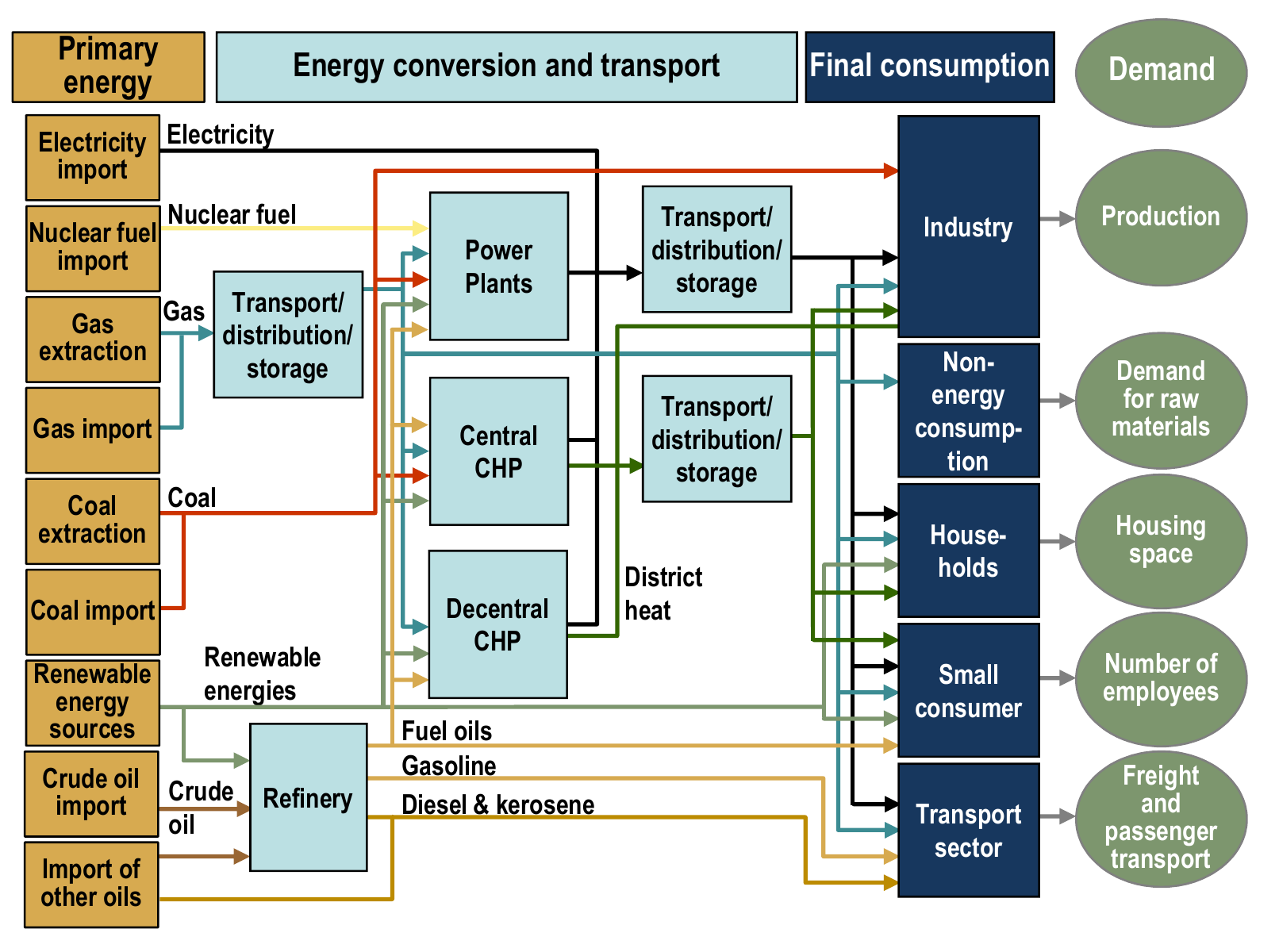}
\caption{\label{fig:IKARUS}
{Schematic representation of the structure of the integrated energy system model IKARUS.} The model covers the entire process chain from primary energy carriers to final energy demand, covering various sectors with a high technological resolution.
}
\end{figure}

We utilize the energy system model IKARUS \cite{Mart06,LINSSEN20172245,17intESM} which depicts the whole German energy system ranging from primary energy supply across conversion and transport of energy carriers to final energy demand in a technology rich way (some thousand technologies), cf.~Figure \ref{fig:IKARUS}. The underlying linear optimization model consists of energy and material flow balances complemented by constraints and further user defined equations. Typically, it is applied to long-term time horizons (currently up to 2050). IKARUS represents Germany as one region and inter-annual variations are included with representative time slices combined with basic heuristics for backup needs in order to guarantee the security of supply.

We chose a current policies scenario framework for all cases analysed here, which takes all already decided policies into account. This comprises especially several energy related laws in Germany, like the German Energy Saving Ordinance (EnEV) or the Renewable Energy Sources Act (EEG). A more detailed description of the included legally binding constraints can be found in \cite{HEINRICHS2017234, Heinrichs2017-eem}. No political intentions or goals are taken into account in this type of scenario framework. Fuel prices are assumed in accordance to the World Energy Outlook 2016 (450ppm scenario) \cite{WEO16}, but as we apply our model for this analysis only to 2020, impacts of fuel price pathways are limited. The electricity exchange is fixed exogenously to avoid mixing effects in the obtained results for this analysis. In addition we assume that the CO$_2$ price is valid for the whole energy system in contrast to the current EU emission trading system in order to get around possible inter-sectoral effects between EU-ETS and Non-EU-ETS sectors and to allow to identify the counter-intuitive effects more clearly.

The fixed budget for case B was derived by minimizing the system costs within IKARUS assuming a CO$_2$ price of 50 Euro/t. Hence, no feasible solution can be found above 50 Euro/t. This value was chosen as it is substantially higher than the current price level of below 10 Euro/t \cite{EEX_CO2}.

The year 2020 was chosen as (i) it allows us to use a current policies scenario, (ii) there will still be a substantial share of fossil based fuels left in the energy system and (iii) it is before the complete nuclear phase-out in Germany (2022) such that mixing effects can be avoided.

\section{Results}

\subsection{Fundamental Model}

We first introduce the basic setup and phenomena for the elementary decision problem described in section \ref{sec:el-mod} before turning to more complex models. Suppose that a country uses two types of fossil fuels for electricity generation: one cheap type with high specific CO$_2$ emissions (e.g. lignite) and one expensive type with low specific CO$_2$ emissions (e.g. natural gas). What is the optimal operation of this electricity system with respect to the three conflicting targets reliability, affordability and reduction of CO$_2$ emissions (cf.~Fig.~\ref{fig:triangle})?

Techno-economic energy system models typically optimize one of the targets while constraints are imposed to the remaining targets. Probably the most common approach is to minimize the total system costs leading to the optimization problem
\begin{align}
\label{eq:opt-costs}
   \mbox{case A:}  \quad
  &  \min_{G_{\rm lignite},G_{\rm gas}} C(G_{\rm lignite},G_{\rm gas}), \\
  & {\rm s.t.} \; E \le E_{\rm cap}, \; G_{\rm gas} + G_{\rm lignite} \ge D,
  \nonumber
\end{align}
with quantities defined in the methods section \ref{sec:el-mod}. Alternatively, one can minimize the total emissions while a budget constraint is applied leading to the optimization problem
\begin{align}
   \label{eq:opt-emissions}
   \mbox{case B:} \quad &
     \min_{G_{\rm lignite},G_{\rm gas}} E(G_{\rm lignite},G_{\rm gas}), \\
    & {\rm s.t.} \; C \le C_{\rm budget}, \; G_{\rm gas} + G_{\rm lignite} \ge D.
    \nonumber
\end{align}
In both cases, the conflict of targets expressed by objectives and constraints can lead to paradoxical effects. This can result in misleading advice for the regulation and governance of the energy system.

\begin{figure}[tb]
\centering
\includegraphics[width=\columnwidth, angle=0]{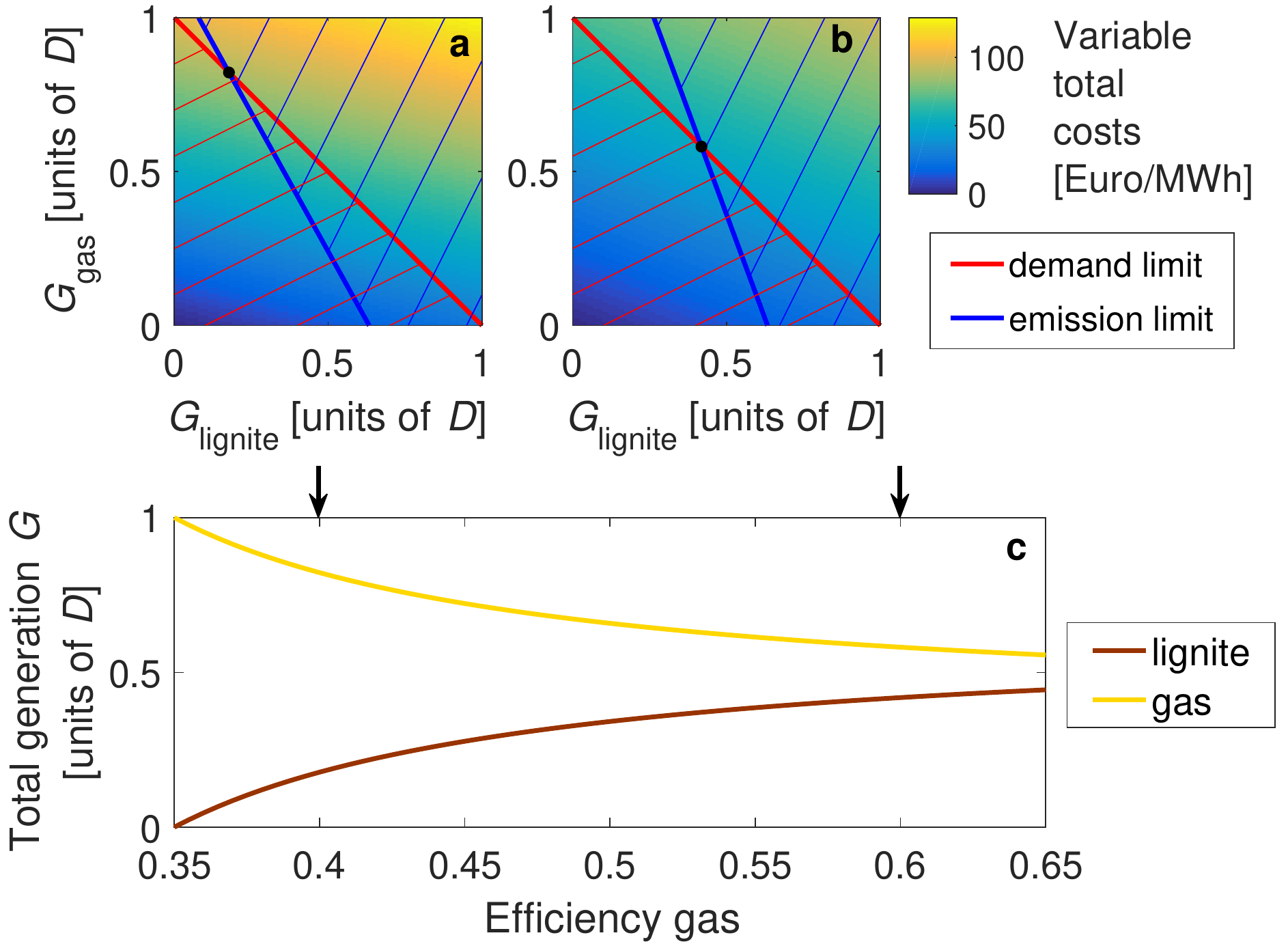}
\caption{\label{fig:element-cost}
{Paradoxical effects occurring in the emission-constrained optimization problem (case A).}
(a,b) Structure of the optimization problem (Eq.~\eqref{eq:opt-costs}) for different values of the efficiency of gas-fired power plants. The objective function is shown in a colour scale, the constraints as thick lines and the infeasible region is striped. When the efficiency of gas-fired power plants increases from 0.4 (a) to 0.6 (b), the emission limit line (blue) moves to the right. This relieves some part of the CO$_2$ budget, which is used for cheap lignite.
(c) An increasing efficiency leads to a smaller utilization of gas-fired power plants.
}
\end{figure}

\begin{figure}[tb]
\centering
\includegraphics[width=\columnwidth, angle=0]{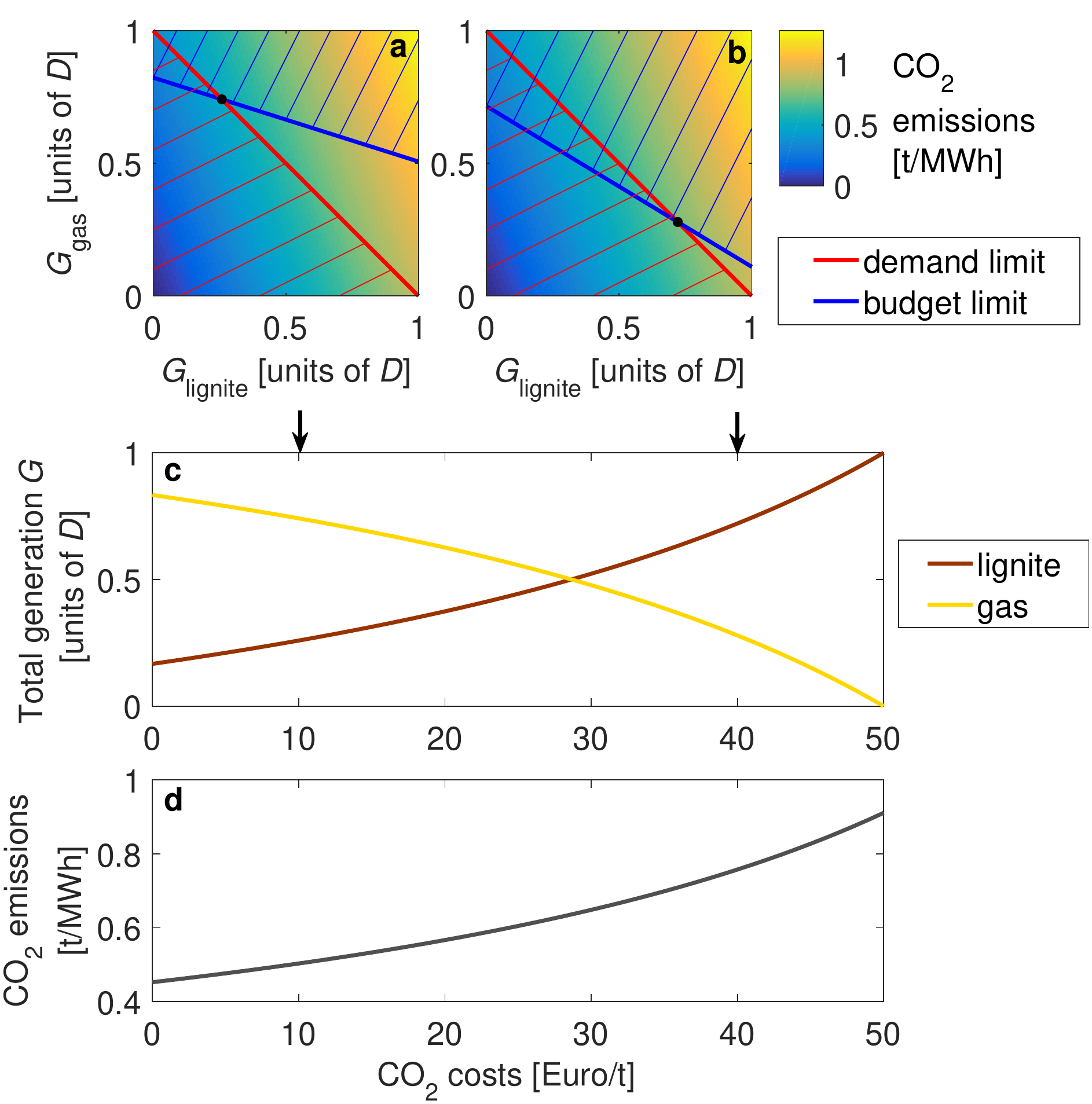}
\caption{\label{fig:element-emission}
{Paradoxical effects occurring in the budget-constrained optimization problem (case B).}
(a,b) Structure of the optimization problem (Eq.~\eqref{eq:opt-emissions}). The objective function is shown in a colour scale, the constraints as thick lines and the infeasible region is striped. When the effective CO$_2$ costs increase from 10 Euro/t (a) to 40 Euro/t (b), the budget limit line (blue) moves downwards such that less money can be spent on the ``cleaner" alternative gas.
(c,d) Increasing CO$_2$ costs thus lead to higher utilization of the cheaper lignite and an increase of CO$_2$ emissions.
}
\end{figure}

Let us first consider case A, whose structure is illustrated in Fig.~\ref{fig:element-cost}a. The emission and reliability constraints exclude many possible combinations of $G_{\rm lignite}$  and $G_{\rm gas}$, leaving only a small feasible region in configuration space. Minimizing costs favors lignite, having low variable costs, over natural gas, having high variable costs. Thus, the optimal solution is found at the right-most point of the feasible region, which is given by the intersection of the demand line and the emission line.

Consider now a technological development, which could allow for emission reductions. Intuitively, one might expect that such a technology is extensively used if the development is cheap enough -- but this expectation can be highly misleading. Assuming that the efficiency of natural gas-fired power plants can be improved \emph{without} any additional costs, we observe a striking effect on the optimal solution shown in Fig.~\ref{fig:element-cost}b. The specific emissions $se_{\rm gas}$ decrease such that the emission line moves to the right. Thus, the intersection point of the emissions and demand lines also moves to the right, resulting in a system optimum that contains \emph{more} lignite and \emph{less} natural gas. The technological improvement of gas-fired power plants essentially frees a fraction of the CO$_2$ cap, which is not used for climate protection but for cost reduction favouring lignite. As a consequence, the share of gas in the energy mix decreases monotonically with the efficiency of gas-fired power plants (Fig.~\ref{fig:element-cost}c). Increasing the efficiency of lignite-fired power plants also relieves the CO$_2$ cap and thus leads to a reduced usage of gas-fired power plants, too.

Constrained optimization can thus lead to a paradoxical effect in energy systems planning: The improvement of a technology may impede its utilization. Such an effect could counter-act incentives for technological innovations and must be compensated by suitable policy measures. In particular, emission caps should be updated either directly or via price-feedback mechanisms. Similar effects are known in cap-and-trade schemes, where technological innovations to reduce emissions simply reduce the price of pollution permits, thus making it cheaper for higher emitters to pollute \cite{Kahn2006}, see also the review \cite{Fawcett2014}. This has led to declines in technological innovation for sulfur dioxide and nitrogen oxide cap-and-trade systems \cite{Taylor2012}.

The paradoxical effects are even more apparent when emissions are minimized in the presence of a budget constraint (case B). Figure \ref{fig:element-emission}a illustrates the structure of the optimization problem -- the system optimum is found at the intersection of the demand line and the budget line. An increase in effective CO$_2$ costs via certificates or taxes \cite{pezzey2013,bertram2015,edenhofer2015,cramton2017global} has a dramatic effect as shown in Fig.~\ref{fig:element-emission}b. The budget line moves downwards such that the intersection of budget and demand line moves to the bottom right in configuration space -- the system optimum contains \textit{more} lignite and \textit{less} gas. In other words: The increase of CO$_2$ costs consumes a part of the restricted budget. In order to compensate for this, expensive natural gas is replaced by cheap lignite.

The paradoxical consequences become most visible in Fig.~\ref{fig:element-emission}c and d. Increasing CO$_2$ costs in the presence of a budget constraint leads to a higher utilization of lignite and thus to higher CO$_2$ emissions. In a real-world setting, such a paradoxical behavior could lead to a complete misdirection of the energy transition -- but only in the presence of a strict budget constraint. In conclusion, a regulatory setting as described here should definitely be avoided (see Discussion).

The observed effect is reminiscent of Giffen's paradox in microeconomics \cite{Stigler1947,Heijman2002}: The demand for an inferior good (here: lignite) \emph{increases} with increasing prices. This is in sharp contrast to our everyday observations formalized in the law of demand that holds for normal goods. Indeed, empirical evidence for a Giffen-type behaviour in real markets has been strongly debated so far (see, e.g. \cite{Stigler1947,Jensen2008}).

\begin{figure*}[tb]
\centering
\includegraphics[width=15cm, angle=0]{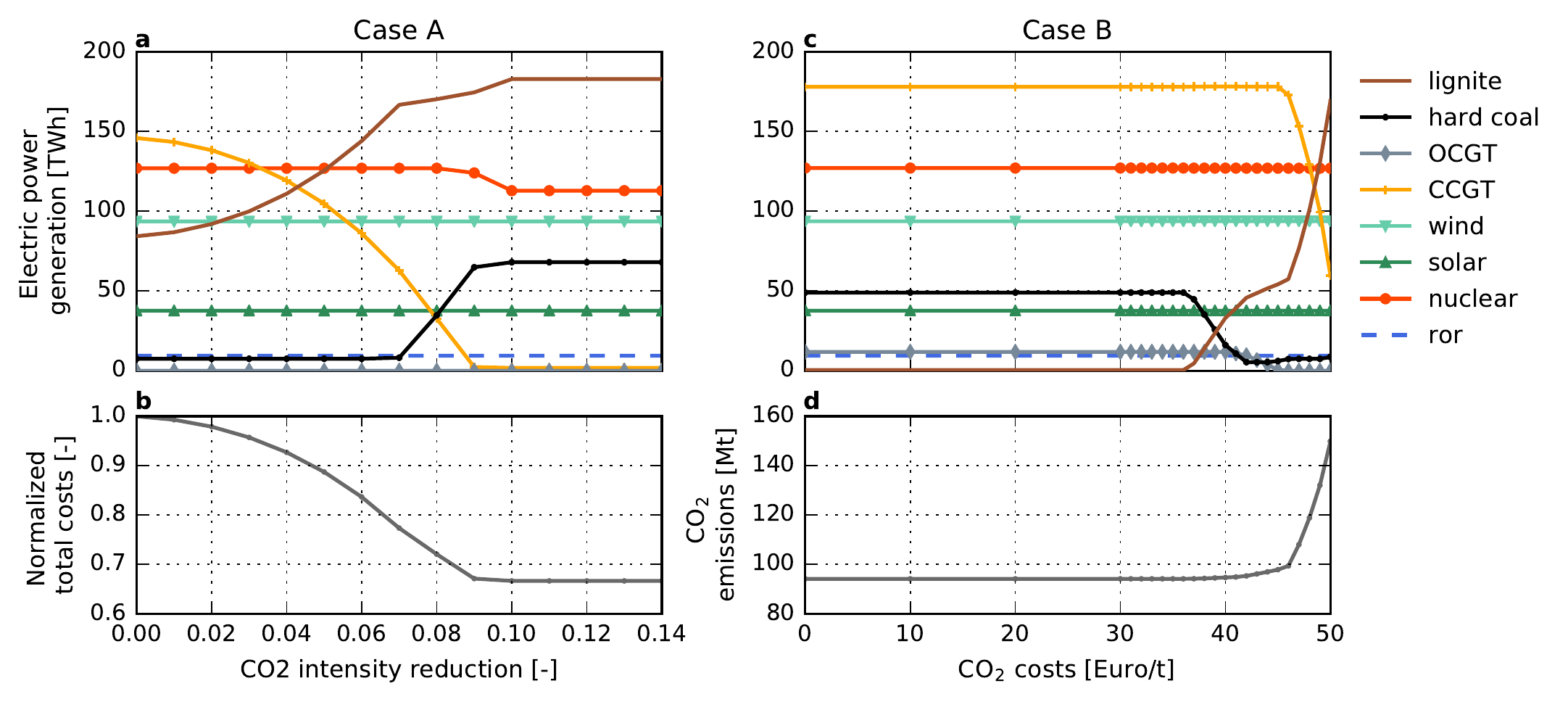}
\caption{\label{fig:pypsa_results}
{Paradoxical effects obtained using the electricity system model PyPSA.}
(a,b) Results for the emission-constrained optimization problem (case A). We analyse the impact of a reduction of the CO$_2$ intensities for each CO$_2$-emitting technology (e.g. via CCS). A CO$_2$ intensity reduction of 0.1 means that the CO$_2$ emissions decrease by 10\% per MWh$_{\rm th}$ for each technology.
(c,d) Results for the budget-constrained optimization problem (case B). We analyse the impact of increasing CO$_2$ costs on the system optimum.
OCGT: open cycle gas turbines, CCGT: combined cycle gas turbines, ror: run-of-river.
}
\end{figure*}

\subsection{Electricity Sector Model}

The paradoxical effects introduced above become manifest in realistic energy system models. We first consider the electricity sector model PyPSA introduced in section \ref{sec:pypsa}. The operation of conventional generators, storage facilities and curtailment of renewable power sources is optimized given a time series for the electricity demand and renewable power availability, respecting power grid reliability constraints (see methods section \ref{sec:pypsa} for details).

We first consider the minimization of total system costs in the presence of a strict cap for CO$_2$ emissions (case A). The system optimum is shown in Fig.~\ref{fig:pypsa_results}a and b as a function of the ``CO$_2$ intensity reduction'', where an intensity reduction of 0.1 corresponds to 10 \% less CO$_2$ emissions per MWh$_{\rm th}$ for each technology, spread across the generation fleet. 

We find that the utilization of natural gas (combined cycle gas turbine, CCGT) rapidly drops to zero when the CO$_2$ intensity is reduced. First, gas is replaced by lignite, then by hard coal, until it is replaced completely for CO$_2$ intensity reductions of approximately 9\%. Remarkably, the share of nuclear energy also decreases slightly.

The detailed simulation of the electricity system thus confirms the paradoxical results outlined above: A reduction in the CO$_2$ intensity frees a fraction of the emission cap, which is not used for climate protection, but for cost reduction. The minimization of system costs always favors fuels with lower variable costs (see Table~\ref{tab:pypsa_costs} for the PyPSA costs), such that lignite and hard coal replace natural gas in the energy mix. Renewables, which have no variable costs, are not affected at all. In conclusion, a technological development reducing CO$_2$ emissions can have quite unintended consequences in the electricity system.

We further simulate case B, where the total emissions are minimized in the presence of a fixed budget constraint. Optimization results are shown in Fig.~\ref{fig:pypsa_results}c and d as a function of effective CO$_2$ costs, implemented via taxes or certificates. We find that moderate CO$_2$ costs up to 36 Euro/t have no effect on the system optimum. Increasing the costs further introduces a dramatic shift in the energy mix. First, cheap lignite replaces hard coal, then open cycle gas turbines (OCGT) and finally combined cycle gas turbines (CCGT) to meet the budget limit. For CO$_2$ costs above 50 Euro/t the optimization problem becomes infeasible by model design -- no solution can be found that comprises the given budget limit. As the variable costs for renewables are zero, they are not affected here. We thus confirm the Giffen-like behaviour outlined above: The increase of CO$_2$ costs in the presence of a budget limit leads to a higher utilization of CO$_2$-intense technologies.

Our results reveal the decisive role of constraints in energy systems optimization. Technological improvements in the presence of a CO$_2$ cap (case A) can lead to a higher utilization of inferior technologies (such as lignite) if constraints (here: the CO$_2$ cap) are not adjusted adequately. In contrast, when minimizing the CO$_2$ emissions in the presence of a budget constraint (case B), increasing CO$_2$ costs may paradoxically lead to an increase of actual CO$_2$ emissions. This is because the system optimum is mainly determined by the budget and demand constraints and only a small feasible region is left in configuration space to minimize the objective function.

\subsection{Integrated Energy System Model}

Paradoxical effects are further analysed using the integrated energy system model IKARUS introduced in section \ref{sec:ikarus}. IKARUS is specially adapted to study the impact of national policy measures affecting the whole energy system such as budget limits, ${\rm CO}_2$ taxes or certificates. While the effects of strict pollutant caps have been discussed before \cite{Kahn2006,Taylor2012}, we here focus on minimizing overall CO$_2$ emissions with a given budget constraint (case B) and study cross-sectoral impacts of increasing effective CO$_2$ prices in the following. We assume that these prices apply to the entire energy system, not just the electricity sector. 

\begin{figure*}[tb]
\centering
\includegraphics[trim={0mm 0mm 5mm 6mm},clip,width=15cm, angle=0]{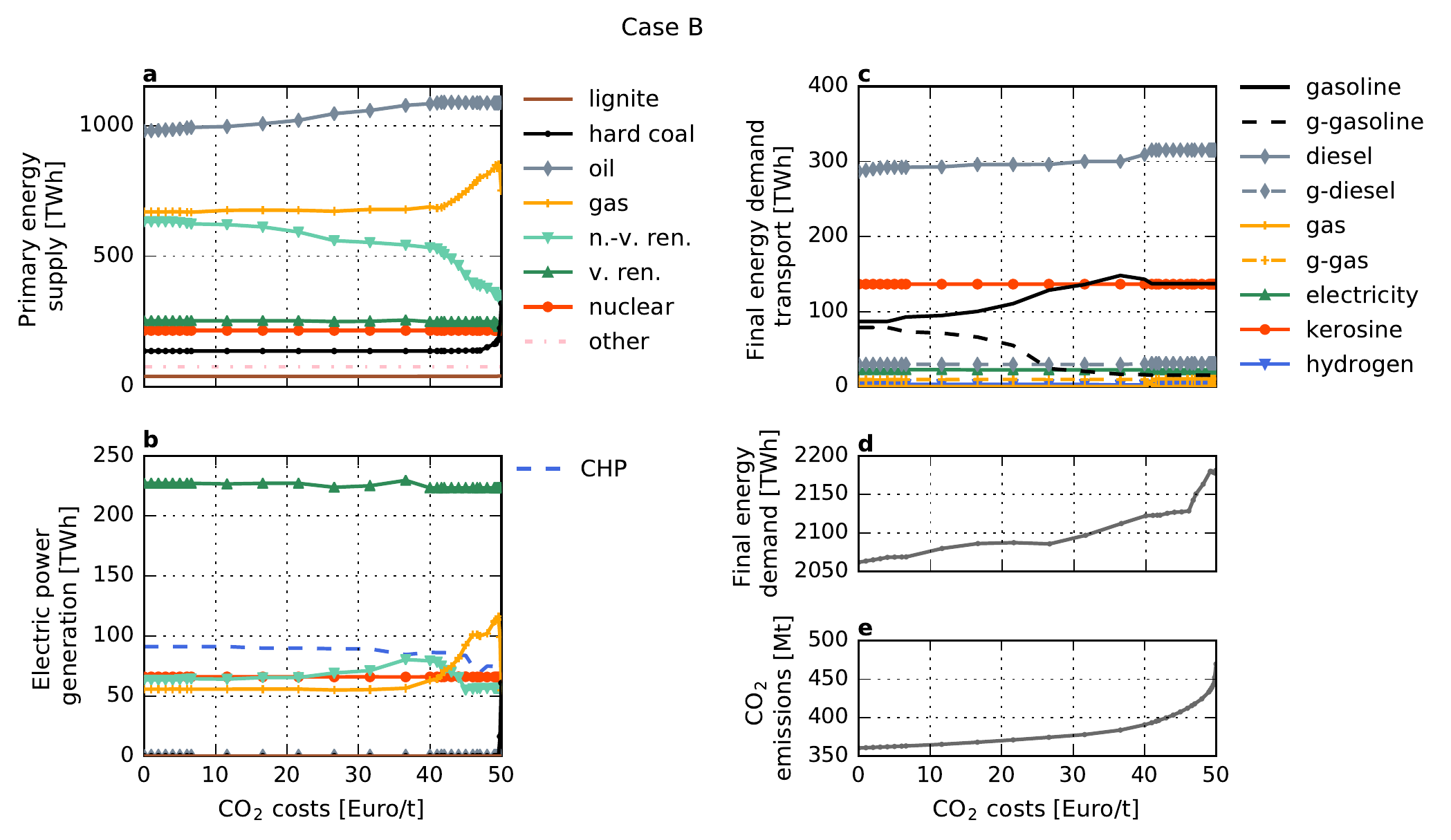}
\caption{\label{fig:ikarus_results}
{Paradoxical effects in the integrated energy system model IKARUS.} We consider the optimization of total CO$_2$ emissions in the presence of a strict budget constraint (case B).
(a) An increase of the effective CO$_2$ price induces a shift from high-cost low-emission to low-cost high-emission energy carriers such as gas, oil and eventually coal, replacing in particular non-variable renewable energy sources (biomass).  
(b,c) In the electricity sector, gas replaces biomass and CHP and in the transport sector, conventional gasoline replaces renewable gasoline (referred to green or g-gasoline, respectively).
(d,e) As a consequence, the final energy demand and the total CO$_2$ emissions \emph{increase}. 
}
\end{figure*}

It is found that increasing CO$_2$ prices leads to a {\em decreasing} usage of non-volatile renewables (here: biomass) in the primary energy supply, which are replaced by energy carriers with higher specific emissions such as gas, oil and eventually coal throughout different sectors (Fig.~\ref{fig:ikarus_results}a). In the transport sector, synthetic fuels are replaced by conventional fossil fuels already for moderate CO$_2$ prices (Fig.~\ref{fig:ikarus_results}b). The electricity sector shows a row of shifts: first from combined heat-and-power (CHP) plants to non-volatile renewables, then from non-volatile renewables to natural gas, and finally from gas to hard coal. Notably, CHP power plants also change their fuel mix with increasing CO$_2$ prices: They are predominantly fuelled by biomass and waste at lower prices and by waste, natural gas and hard coal at higher prices. The usage of volatile renewable energy sources and nuclear energy as well as process based fuel demand (e.g.~due to steam demand of industry processes) are not significantly affected by CO$_2$ prices, because they have no CO$_2$ emissions or cannot be substituted easily (e.g. the steam demand in industry).

In addition to the usage of energy carriers, the efficiency of technologies plays a major role in the integrated energy system model IKARUS. CHP is a prime example as it enhances the efficient use of primary energy carriers by providing heat and electric power simultaneously. With increasing CO$_2$ prices no budget is left for the implementation of expensive efficiency measures. Less efficient technologies have to be used such that total CO$_2$ emissions increase (Fig.~\ref{fig:ikarus_results}e).  

Finally, both the substitution and the efficiency effects contribute to an increase in the final energy consumption and the total CO$_2$ emissions with increasing CO$_2$ prices (Fig.~\ref{fig:ikarus_results}d and e) as reported before. Hence, our results confirm the decisive role of constraints in energy systems optimization revealed by the previous described elementary and electricity sector models.

\section{Discussion}

The realization of a sustainable, affordable and reliable energy supply is a major technological and political challenge. Techno-economic optimization models are central tools for the planning and analysis of future energy systems. Frequently, different energy policy targets are translated into one objective function and several constraints. In this paper, we have demonstrated several surprising optimization results arising from the interactions between these targets.

When system costs are minimized in the presence of a ${\rm CO}_2$ cap, efficiency gains may have counter-intuitive effects. An increase in efficiency frees a part of the ${\rm CO}_2$ cap, allowing cheap high-emission technologies to replace low-emission technologies. This mechanism could impede incentives for a technological development such that an adequate adjustment of the CO$_2$ cap is be needed. These aspects should be kept in mind in the discussion about appropriate policy measures to reduce emissions, in particular in the decision for cap-and-trades versus taxes.

Even more striking results are observed when emissions are minimized in the presence of a budget constraint. Increasing  CO$_2$ prices can oust clean, but expensive technologies out of the system, and eventually lead to higher emissions. To our knowledge such a fixed budget is currently not implemented in any free energy market, but some forms of caps have been repeatedly demanded in political discussions: Energy price caps have been suggested in Austria \cite{ViceAT2003,PresidentATInd2004b} and the United Kingdom \cite{EdMili2013}, where they even appeared in the governing Conservative party's 2017 election manifesto \cite{ConsManifesto17} and are likely to be implemented by the end of 2018 \cite{UKParliament18} (see appendix for exact quotes). Our results show that such a budget limit is incompatible with ${\rm CO}_2$ taxes or certificates, in the sense that the interference of both measures can drive the energy transition into an unwanted direction. Hence, more effective regulatory means should be applied.

We conclude that extreme care is necessary in the design and interpretation of energy systems optimization models as every constraint can have a decisive impact on the result. The transparency of such models must be improved \cite{PFENNINGER201863,Pfenninger2017,PFENNINGER2017b} and the complex interactions of different regulatory measures implemented via caps or prices must be thoroughly analysed and respected in any policy advice.  

\section*{Appendix}

Here we give more details on the political claims for a (partial) limitation of the energy budget or electricity prices mentioned in the discussion:

Retail electricity prices were capped in California for several years, which contributed to the 2000/01 Western US Energy Crisis \cite{Sweeney2013}.

On the 26.11.2003, the vice president of the Wirtschaftskammer \"Osterreich (Austrian chamber of commerce) claimed a "Deckelung der Energiekosten in der Industrie". The press release is available on the website of the Austrian Press agency APA at \cite{ViceAT2003}.

The president of the Austrian Industriellenvereinigung (Industry Association) claimed during a press conference on 8.11.2004 "das Einziehen eines Gesamtdeckels f\"ur Energiekosten - inklusive Kosten f\"ur den Emissionshandel und der Energiesteuer - sowie der Ausbau des Leitungsnetzes und der Stromerzeugung in \"Osterreich." The original quote and a further discussion are available from the website of the Austrian Press agency APA at \cite{PresidentATInd2004b}.

In 2013, the head of the British labour party Ed Miliband said that "Labour would freeze gas and electricity bills for every home and business in the UK for 20 months if it wins the 2015 election". The quote and the original video are available from the website of the British Broadcasting Corporation BBC, dated 24.9.2013, available online at \cite{EdMili2013}.

In 2014, the prime minister of the German federal state Bavaria, Horst Seehofer, proposed an upper limit for the subsidies for renewable energy sources at eight Euro Cent per kWh: "Etwa acht Cent w\"are eine Zahl, \"uber die man mal reden muss". The original quotes and a further discussion are available from the website of the newspaper Sueddeutsche Zeitung, dated 11.3.2014, online at \cite{Seehofer2014}.

The British conservative party proposed an absolute tariff cap in its 2017 election manifesto: "We will introduce a safeguard tariff cap that will extend the price protection currently  in  place  for  some  vulnerable  customers  to  more  customers  on  the  poorest  value tariffs" \cite{ConsManifesto17}. The initiative of the Conservative party is supported by the British Parliament and thus likely to be implemented. A pre-legislative scrutiny of the draft Domestic Gas and Electricity (Tariff Cap) Bill by a committee of the House of Commons concludes: "These  repeated  failures  have  led  us  to  support  the  Government's  initiative  to  set  a  temporary absolute price cap on standard variable and default tariffs" \cite{UKParliament18}.

\section*{Acknowledgments}

We gratefully acknowledge support from the Helmholtz Association via the joint initiative ``Energy System 2050 -- A Contribution of the Research Field Energy'', the grant no.~VH-NG-1025 to D.W. and grant no.~VH-NG-1352 to T.B.


\end{document}